\documentclass[preprint, secnumarabic, amssymb, nobibnotes, aps]{revtex4-2}

\usepackage[T1]{fontenc}

\usepackage{geometry}
\usepackage{setspace}

\usepackage{graphicx}
\usepackage{float}
\newfloat{scheme}{htbp}{los}
\floatname{scheme}{Scheme}
\floatname{chart}{Chart}
\newfloat{graph}{htbp}{loh}
\usepackage{ulem}
\usepackage{chemformula} 
\usepackage[version = 4]{mhchem} 

\usepackage{xcolor}

\newcommand{\ie}[1]{{\it i.e.}}
\newcommand{\eg}[1]{{\it e.g.}}
\usepackage{graphicx}
 \usepackage{amsmath}
 \usepackage{array}
 \usepackage{physics}
\usepackage{ulem}

\begin{document}

\title{Determination of the Fermi Energy of Diamond using Photoluminescence Spectral Analysis}

\author{Yifan Song}%
\affiliation{Department of Chemistry, University of Southern California, Los Angeles, CA 90089, USA}

\author{Sina Ilkhani}
\affiliation{Department of Chemistry, University of Southern California, Los Angeles, CA 90089, USA}

\author{Leah Webb}
\affiliation{Department of Chemistry, University of Southern California, Los Angeles, CA 90089, USA}

\author{Derrick Lin}
\affiliation{Department of Chemistry, University of Southern California, Los Angeles, CA 90089, USA}

\author{Gem Nakamura}
\affiliation{Department of Physics \& Astronomy, University of Southern California, Los Angeles, CA 90089, USA}

\author{Helen Highland}
\affiliation{Department of Chemistry, University of Southern California, Los Angeles, CA 90089, USA}

\author{Stephen B. Cronin}
\email[Corresponding email: ]{scronin@usc.edu}
\affiliation{Ming Hsieh Department of Electrical Engineering, University of Southern California, Los Angeles, CA 90089, USA}
\affiliation{Department of Chemistry, University of Southern California, Los Angeles, CA 90089, USA}
\affiliation{Department of Physics \& Astronomy, University of Southern California, Los Angeles, CA 90089, USA}

\author{Susumu Takahashi}%
\email[Corresponding email: ]{susumuta@usc.edu}
\affiliation{Department of Chemistry, University of Southern California, Los Angeles, CA 90089, USA}
\affiliation{Department of Physics \& Astronomy, University of Southern California, Los Angeles, CA 90089, USA}
\affiliation{Center for Quantum Information Science and Technology (CQIST), University of Southern California, Los Angeles, CA 90089, USA}
\date{\today}%


\begin{abstract}
Electronic band structures and the Fermi energy provide essential information for understanding the electronic properties of solids. In semiconductors, the Fermi energy is determined by the donor and acceptor concentrations. For diamond, the relationship between the Fermi energy and the donor-acceptor concentrations is highly nonlinear; therefore, experimental determination of the Fermi energy is important. 
Here, we report a method to determine the Fermi energy of diamond based on photoluminescence (PL) measurements. 
The density-functional-theory (DFT) study by Deák et al.~\cite{deak2014formation} showed the relationship between the Fermi energy and the formation energies of nitrogen-vacancy centers in the negatively charged (NV$^-$) and neutrally charged (NV$^0$) charge states.
In the present method, we measure the relative populations of the NV$^-$ and NV$^0$ centers from PL spectral analysis and determine the Fermi energy using these populations and the DFT result.
Moreover, we show the application of the method by extracting the Fermi energies of several diamond samples and discuss the spin coherence times of their NV centers and their stability against the charge state conversion. 
The PL-based method is advantageous for determining the Fermi energy with high spatial and fast time resolutions, even in extreme environments, and can be extended to determine various wide bandgap semiconductors.
\end{abstract}

\maketitle

\section{Introduction}
In solid state physics, the electronic band structures and the Fermi energy of the system provide essential information to understand the electronic, optical, and spin properties of materials. For example, the metallic properties of a solid can be understood based on the possession of bands within which the Fermi energy resides. The conducting property of a semiconductor can also be understood by the level of the Fermi energy within the band gap, which is determined by the amounts of donors and acceptors in the material~\cite{Ashcroft76,kittel2018introduction,marder2010condensed}.

Diamond is a semiconductor with a band gap of 5.5 $eV$. 
Defects in diamond, such as nitrogen-vacancy (NV) and silicon-vacancy (SiV) centers, have recently attracted significant attention due to their unique spin and optical properties and their potential applications in quantum science and technology~\cite{doherty2013nitrogen, schirhagl2014nitrogen, maze2008nanoscale, balasubramanian2008nanoscale, taylor2008high, li2021determination, fortman2020demonstration, ren2023demonstration, bersin2024telecom, stas2022robust}.
For diamond crystals, the common donors and acceptors are nitrogen and boron atoms, respectively. Therefore, the Fermi energy is often located between 0.4 $eV$ and 3.8 $eV$ from the valence band, corresponding to the levels of the boron acceptor and nitrogen donor, respectively~\cite{collins2002fermi, gali2013ab}. Because the energy levels of the donors and acceptors are relatively deep to the thermal energy of the room temperature, the dopants are hardly ionized at room temperature. As a result, the concentration dependence on the Fermi energy is highly non-linear~\cite{collins2002fermi}. 
Therefore, an experimental method to measure the Fermi energy is extremely important for the engineering of diamond properties.

In this paper, we present a method for non-destructively, instantaneously, and robustly determining the Fermi energies of diamond crystals. The method is based on photoluminescence (PL) measurements.
It is known that the defect centers in diamond, such as the NV and SiV centers, exhibit multiple charge states. For instance, while most of the investigations on the NV center are targeted at the negatively charged NV$^{-}$ center, it is also known that the NV$^{-}$ center undergoes charge
state conversion to the neutrally charged NV$^{0}$ center under laser excitation~\cite{aslam2013photo, shields2015efficient}.
Moreover, density functional theory (DFT) studies show that the NV center can be stabilized in doubly negative (NV$^{2-}$), negative (NV$^{-}$), neutral (NV$^{0}$), and positive (NV$^{+}$) charge states, and the stable charge state of impurity centers in diamond is determined by the level of the Fermi energy~\cite{gali2013ab, deak2014formation, garcia2024photo, thiering2018ab}. 
In the present study, we utilize the combination of PL spectrum analysis and
DFT result to obtain the Fermi energy.
First, using the PL spectral analysis of the NV$^{-}$ and NV$^{0}$ centers, we extract their relative populations. In the relative-population analysis, we study the laser-power dependence to obtain the population ratio in the absence of laser excitation. After obtaining the population ratio, we use a DFT-derived model that relates the population ratio to the Fermi energy. 
The present method is applied to determine the Fermi energies of eight diamond samples, including three nitrogen-doped and five thermal-grade diamond crystals.
We also discuss the correlation between the Fermi energy, the spin decoherece time ($T_2$), and the stability against the charge state conversion.
This optical measurement approach provides fine spatial and fast temporal resolution and can operate under a wide range of temperatures, pressures, and electromagnetic fields. 
The study offers new insights into engineering the Fermi energy to control defect charge states and design material properties for quantum technologies.

\section{Materials and Methods}
Diamond samples were investigated using photoluminescence (PL), optically detected magnetic resonance (ODMR), and high-frequency electron paramagnetic resonance (HF EPR) measurements. The samples and experimental details are described below. 
\subsection{Diamond samples}
In this work, we study eight diamond crystals. The crystal samples include three nitrogen-doped diamond (Sample 1-3) and five thermal-grade diamond crystals (Sample 4-8).
Sample 1 and 2 are single-crystal, quantum-grade CVD diamond crystals purchased from Element Six. Sample 1 is DNV14 and Sample 2 is DNV1. The size of Sample 1 and 2 is 3 $\times$ 3 $\times$ 0.5 mm$^3$.
Sample 3 is a high-pressure, high-temperature (HPHT) single-crystal type-Ib diamond purchased from Sumitomo Electric Industries. The size is 2 $\times$ 2 $\times$ 0.3 mm$^3$.
Sample 3 was subjected to successive irradiation using a high-energy (4 MeV) electron beam and annealing processes (at 1000 $^o$C) in order to increase the NV center density. 
Major impurities in Sample 1-3 are nitrogen-related impurities such as single nitrogen substitutional defects (P1 centers) and NV centers.
According to Element Six, DNV14 (Sample 1) has the P1 and NV concentrations of 13 and 4.5 ppm, respectively. DNV1 (Sample 2) has the P1 and NV concentrations of 800 and 300 ppb, respectively.
For Sample 3, the P1 and NV concentrations were estimated to be more than 12.5 and 1 ppm, respectively~\cite{fortman2020demonstration}.
In addition, Sample 4-8 are thermal-grade polycrystalline CVD diamond (TM220 and TM200) purchased from Element Six. The thermal-grade diamond is developed as a thermal management material for high-power RF, opto-electronics, and high-voltage power semiconductor devices. According to the description in Element Six, TM220 has a higher thermal conductivity value than TM200. Among the thermal grade diamond crystals, two samples are TM220 with a thickness of 500~$\mu$m (indicated as Sample 4 and 5). Sample 6 is also TM220, but the sample was polished and thinned down to a thickness of 170~$\mu$m. Samples 7 and 8 are TM200 with a thickness of 300~$\mu$m. Those thermal grade samples are 10~mm wide and 10~mm long.

\subsection{Photoluminescence}
PL spectra were collected using the Horiba and the Teledyne (HRS-300) spectrometers. In the Horiba system, a 10$\times$ objective lens with a numerical aperture of 0.25 and an 1800 groove/mm grating is used. A 532 nm continuous-wave laser serves as the excitation source. The maximum laser power is 100 mW. In addition, the system has an option to use a 635 nm excitation laser with an output power of 200 mW. The emitted light was captured through the same objective and detected within the spectrometer. 
In the Teledyne system, the 532 nm laser with an output power of 100 mW was used. The system uses a 10$\times$ objective lens with a numerical aperture of 0.25.
The details of the Teledyne system are described elsewhere~\cite{Highland26}.

\subsection{Optically detected magnetic resonance}
The $T_2$ measurements of the NV center were performed by optically detected magnetic resonance (ODMR) spectroscopy. The ODMR spectroscopy was performed using a confocal microscope system and a microwave access, consisting of a continuous-wave diode-pumped solid-state laser (Crystalaser), an acousto-optic modulator (AOM; Isomet), a microscope objective (Nikon), avalanche photo diodes (APDs; Pacer), a XYZ piezo stage (Thorlabs), a microwave source (Stanford Research), and optics~\cite{Fortman19}. 
The 532 nm laser is focused into a diffraction-limited spot using the microscope objective. PL signals of the NV center are collected by the same microscope objective, and then isolated from the laser excitation using a dichroic mirror and a fluorescence filter. Finally, the PL is detected by the APD detector. AOM is used for pulsed ODMR experiments. The sample position is controlled by the XYZ piezo stage. For ODMR spectroscopy, a microwave excitation generated by the microwave source is applied to the NV center through the microwave transmission line (a small gold wire) placed on the diamond surface. No magnetic field is applied. Timing for pulsed ODMR is controlled by a pulse generator (SpinCore). The $T_2$ measurements are carried out using a spin echo sequence of $\pi$/2-$\tau$-$\pi$-$\tau$-$\pi$/2, where we increment the evolution time ($\tau$) to observe the $T_2$ decay behavior, and the last $\pi$/2-pulse is used to map the echo intensity to the PL intensity.

\section{Results and Discussion}
\subsection{NV charge states}
\begin{figure*}[t]
    \centering
     \includegraphics[width=\textwidth]{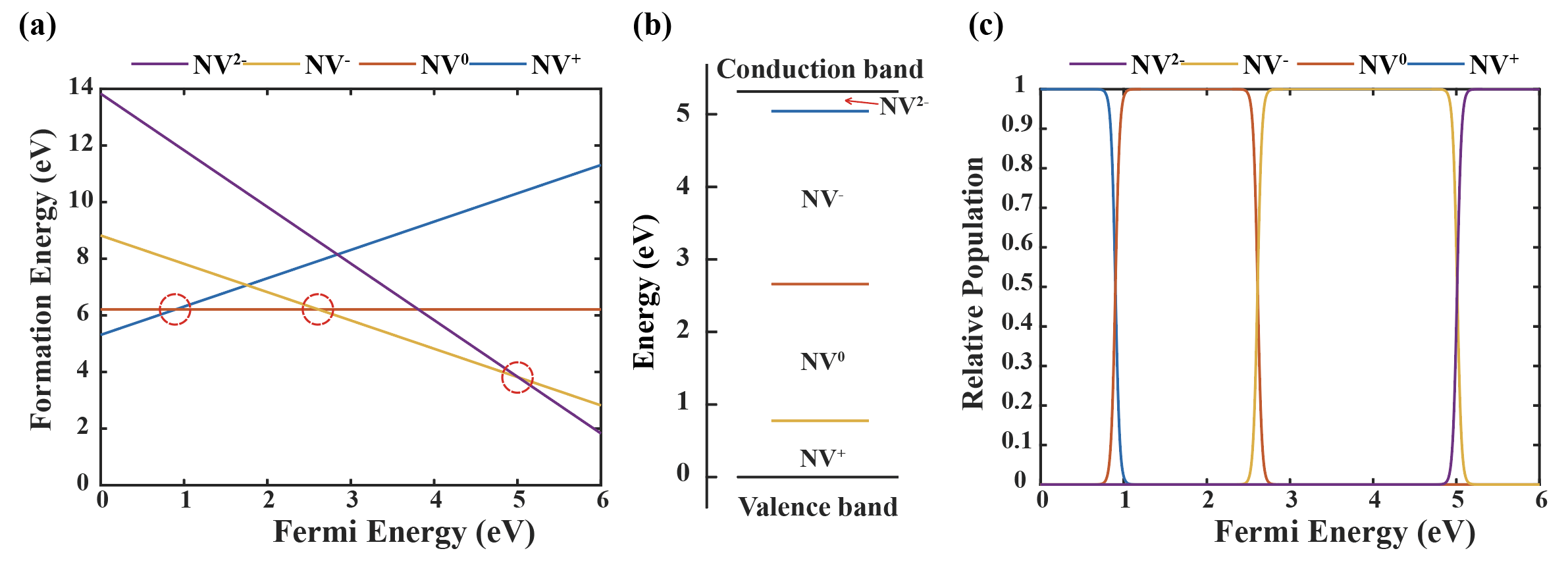}
    \caption{NV charge states and the Fermi energy. (a) The DFT calculation of the formation energy of the NV center as a function of Fermi energy~\cite{deak2014formation}. 
    (b) The energy diagram for different NV charge states with respect to the conduction band and valence band.(c) The relative populations of NV charge states as a function of Fermi energy.}
    \label{fig:formation}   
\end{figure*}
The NV center in diamond can be stabilized in several different charge states. For example, it is well-known that the negatively charged NV (NV${^-}$) center can be converted to the neutrally charged NV (NV$^{0}$) center through laser excitation. The DFT study by Deák et al.~\cite{deak2014formation} shows that the formation energies of the NV charge states are given as follows,
\begin{equation}
\begin{split}
E^{+}_{form}(E_F) = 5.31+E_F \\
E^{0}_{form}(E_F) = 6.21 \\
E^{-}_{form}(E_F) = 8.82-E_F\\
E^{2-}_{form}(E_F) = 13.83-2E_F
\end{split}
\label{eq:e_form}
\end{equation}
where $E_F$ is the formation energy with respect to the energy of the valence band. Using Eq.~\ref{eq:e_form}, a plot for the formation energies as a function of the Fermi energy is shown in Fig.~\ref{fig:formation}(a).
As can be seen, the energy levels for each charge state depend on the Fermi energy. Because the charge state with the lowest level is energetically favorable, the NV$^{0}$ state is favorable when the Fermi level lies in the range of 0.9-2.6 $eV$ from the valence band. In addition, in the range of 2.6-5 $eV$, the NV$^{-}$ state is favorable. When the Fermi energy is near the mid-gap ($\sim$2.6 $eV$), both the NV$^{-}$ and NV$^{0}$ states are favorable. In addition, the NV$^{2-}$ state is favored with Fermi levels above 5 $eV$. 
From the result of Eq.~\ref{eq:e_form}, one can map energetically favorable states as a function of the Fermi energy, as shown in Fig.~\ref{fig:formation}(b). As can be seen, the NV center can exist as positive (NV$^{+}$), neutral (NV$^{0}$), negative (NV$^{-}$), and doubly-negative (NV$^{2-}$) charge states. However, the NV$^{2-}$ state may be challenging to realize because it becomes stable with $E_F$ > 5 eV, which is higher than the nitrogen donor level of $\sim$3.8 $eV$.

Moreover, the relative population of each charge state ($P_q$ where \textit{q} is the number of charges) can be obtained using the formation energies for each charge state and the Boltzmann distribution, \textit{i.e.},
\begin{equation}
P_q (E_F) = \frac{1}{Z}e^{-\frac{E^{q}_{form}(E_F)}{k_BT}}
\label{eq:boltz}
\end{equation}
and the partition function ($Z$) is given by,
\begin{equation*}
Z = \sum_{q}e^{-\frac{E^{q}_{form}(E_F)}{k_BT}}
\label{eq:partition}
\end{equation*}
where $k_B$ is the Boltzmann constant. $T$ is the temperature in Kelvin. In the present case, the investigation was done at $T = 300$ Kelvin. The calculated relative population is shown in Fig.~\ref{fig:formation}(c). As can be seen, one of the charge states dominates the population in most ranges of $E_F$ except in the regime where two formation energy lines cross (indicated in Fig.~\ref{fig:formation}(b)). In the crossing regime, the two corresponding charge states dominate the populations. Therefore, the measurement of the relative population of two NV’s charge states allows us to determine the Fermi energy of the diamond crystal. For example, if both NV$^{-}$ and NV$^{0}$ are equal, the Fermi energy will be $\sim$2.6 $eV$.

\subsection{Decomposition of PL spectrum}
\begin{figure}
    \centering
     \includegraphics[width=\textwidth]{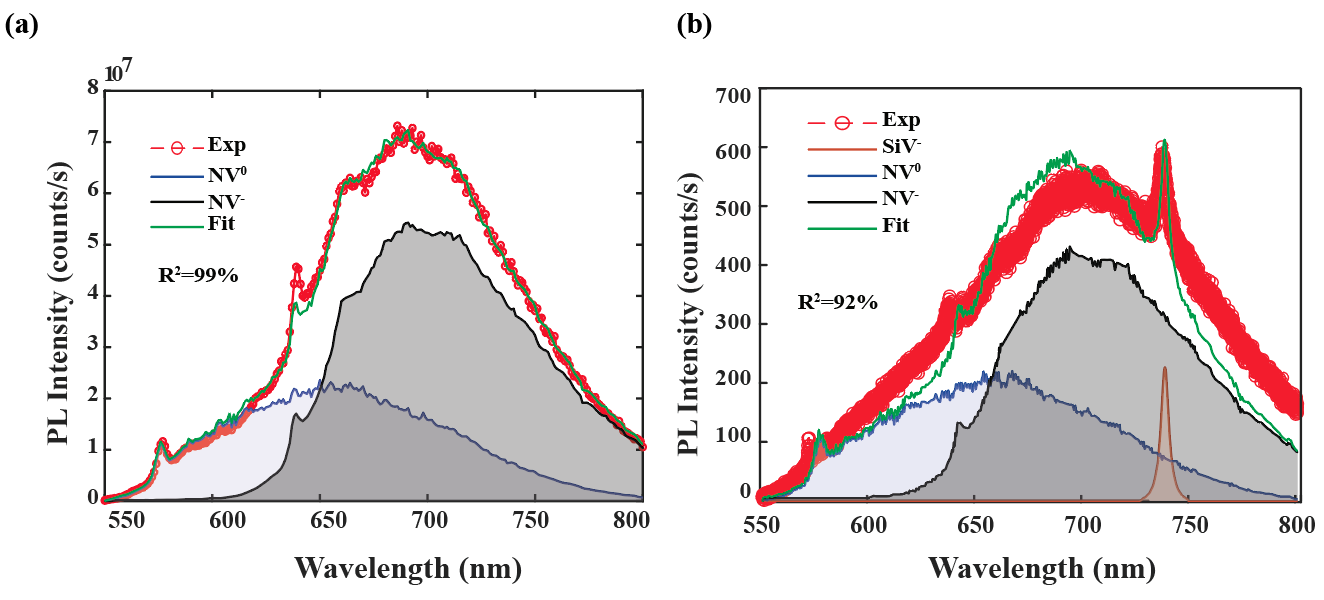}
    \caption{PL spectrum decomposition analysis. 
    (a) PL spectrum of Sample 1. The PL intensity is normalized by the measurement time. The data were taken with a laser power of 14.6 mW and an integration time of 10 ms. The y-axis shows the normalized intensity in units of (count/s). From the fit, we obtained $[NV^{-}]$ and $[NV^{0}]$ to be $0.450 \pm 0.004$ and $0.550 \pm 0.004$, respectively. 
    (b) PL spectrum of Sample 4. The data were taken with a laser power of 10 mW and an integration time of 5 s. The y-axis shows the signal intensity in units of (count/s).  $[NV^{-}]$ and $[NV^{0}]$ are $0.360 \pm 0.002$ and $0.640 \pm 0.002$, respectively. The errors represent the 95\% confidence interval.}
    \label{fig:PLdecomp}   
\end{figure}
Figure~\ref{fig:PLdecomp}(a) shows PL intensity normalized by the measurement time for Sample 1. The PL spectrum was taken with a laser power of 14.6 mW. As can be seen, the observed PL spectrum exhibits a broad signal with a width of more than 200 nm. The spectrum also shows narrow peaks at a wavelength of 637 nm and 575 nm, corresponding to the zero-photon line (ZPL) of the NV$^{-}$ and NV$^{0}$ centers, respectively. Therefore, the spectrum consists of contributions of NV$^{-}$ and NV$^{0}$ centers. 
As discussed previously, the observation of both NV$^{-}$ and NV$^{0}$ centers indicates that the Fermi energy is around 2.6 $eV$, and, in this range of Fermi energy, the populations of the NV$^{+}$ and NV$^{2-}$ centers are negligible. 
Moreover, the PL spectrum of Sample 4 is shown in Fig.~\ref{fig:PLdecomp}(b). The PL spectrum was taken with a laser power of 10 mW. As can be seen, the PL intensity is approximately $10^{-4}$ smaller than that of Sample 1. 
The PL spectrum from Sample 4 also contains only the NV$^{-}$ and NV$^{0}$ contributions. Similar to Sample 1, because both PL signals of NV$^{-}$ and NV$^{0}$ were observed, the populations of the NV$^{+}$ and NV$^{2-}$ centers are also negligible in Sample 4.
In addition, Sample 4 displays a sharp emission peak at 738 nm, attributed to the SiV$^{-}$ center. 

\begin{figure*}[t]
    \centering
     \includegraphics[width=\textwidth]{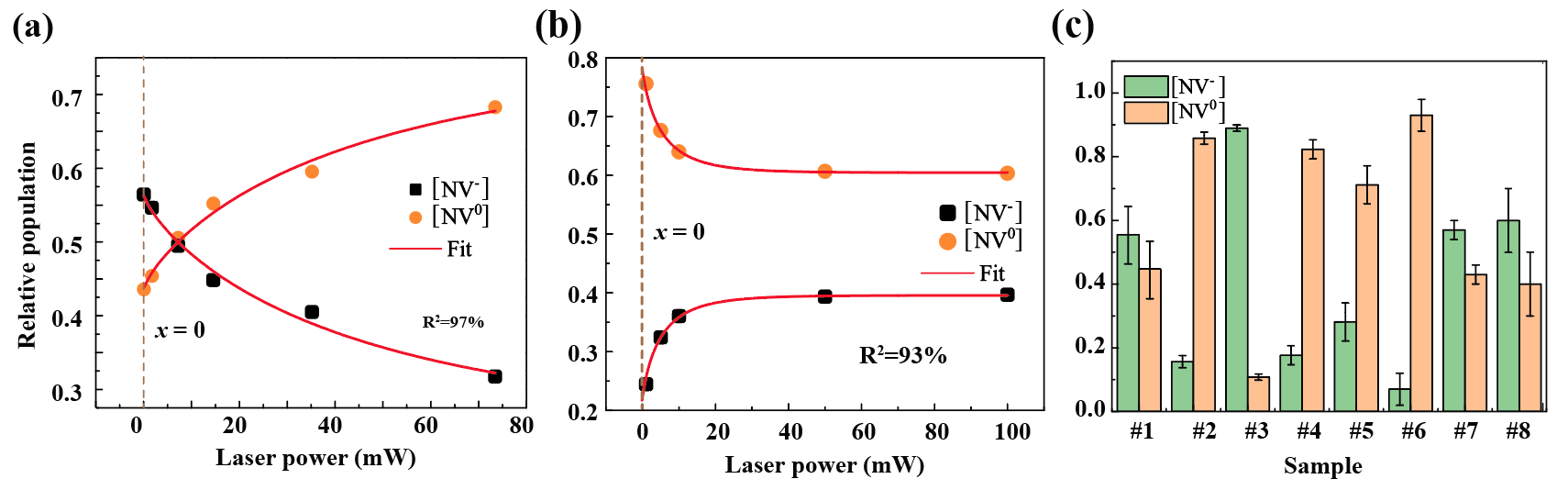}
    \caption{
    $[NV^{-}]$ and $[NV^{0}]$ as a function of laser power. (a)-(b) results from Sample 1 and 4, respectively. Black and orange points: experimental data (square: NV$^{-}$, dot: NV$^{0}$); red line: fitting. The obtained fitting parameters were the following. 
    (a) Sample 1. $[NV^{-}] = y(x)$ and $[NV^{0}] = 1-y(x)$, where $y(x) = y_0+A\exp[-(x/\tau)^n]$. From the fit, $y_0 = 0.25 \pm 0.04$, $A = 0.32 \pm 0.05$, $\tau = 50 \pm 18$, $n = 0.8 \pm 0.0$ were obtained.
    (b) Sample 4. $[NV^{-}] = 1- y(x)$ and $[NV^{0}] = y(x)$. $y_0 = 0.61 \pm 0.01$, $A = 0.21 \pm 0.02$, $\tau = 4.20 \pm 0.80$, $n = 0.70 \pm 0.10 $ were obtained.
    (c) Relative populations for NV$^{-}$ and NV$^{0}$ in the absence of laser excitation ($[NV^{-}]_0$ and $[NV^{0}]_0$).}
    \label{fig:zerolaser}   
\end{figure*}
Next, we employ a decomposition analysis of the PL spectrum to extract the relative populations of the NV$^{-}$ and NV$^{0}$ centers. As reported previously~\cite{pambukhchyan2023probing}, the decomposition analysis enables the quantitative determination of the relative populations of NV$^{-}$ and NV$^{0}$ centers. In the analysis, the PL spectrum was fitted using a linear combination of the standard NV$^{-}$, NV$^{0}$, and SiV$^{-}$ spectra. The standard spectra for the NV$^{-}$ and NV$^{0}$ centers were extracted from the previously reported data~\cite{jeske2017stimulated}. 
The SiV$^{-}$ PL spectrum was modeled by the Lorentzian function. Namely, the total PL spectrum is fitted by,
\begin{equation}
PL_{tot} = c_{NV^-} I_{NV^{-}} + c_{NV^0} I_{NV^{0}} + c_{SiV^-} I_{SiV^{-}}
\label{eq:decomp}
\end{equation}
where $c_{NV^{-}}$, $c_{NV^{0}}$, and $c_{SiV^{-}}$ represent the contributions from NV$^{-}$, NV$^{0}$, and SiV$^{-}$ on the PL spectrum, respectively. $PL_{tot}$ is the total spectrum. $I_i$ is the standard spectrum for $i=$ NV$^{-}$, NV$^{0}$, and SiV$^{-}$. 
Using Eq.~\ref{eq:decomp} and the least-squares regression method, we obtain the best fit results. For samples with no SiV$^{-}$ PL peak, \eg{} Sample 1, we set $c_{SiV^{-}}=0$ in the fit.
Then, $c_{NV^{-}}$ and $c_{NV^{0}}$ are converted to the NV$^{-}$ and NV$^{0}$ populations ($[NV^{-}]$ and $[NV^{0}]$, respectively) using the following equations,
\begin{equation}
\begin{split}
[NV^-]=\frac{c_{NV^{-}}}{c_{NV^{-}}+\kappa_{532}c_{NV^{0}}}\\
[NV^0]=\frac{\kappa_{532}c_{NV^{0}}}{c_{NV^{-}}+\kappa_{532}c_{NV^{0}}}
\end{split}
\label{eq:kappa}
\end{equation}
where $\kappa_{532}$ is the PL emission efficiency ratio between NV$^{-}$ and NV$^{0}$ centers under 532 nm wavelength excitation, and was determined to be 2.5 previously~\cite{alsid2019photoluminescence}. 
As shown in Figs.~\ref{fig:PLdecomp}(a) and (b), the fit demonstrates good agreement with the data. From the result and Eq.~\ref{eq:kappa}, we obtained that $[NV^{-}]$ and $[NV^{0}]$ for Sample 1 are $0.450 \pm 0.004$ and $0.550 \pm 0.004$, respectively. 
The contribution of the SiV$^{-}$ center was ignored since the SiV$^{-}$ PL was not visible. 
In addition, as can be seen from Fig.~\ref{fig:PLdecomp}(b), the fit result for Sample 4 also showed a good agreement, and we obtained that $[NV^{-}]$ and $[NV^{0}]$ are $0.360 \pm 0.002$ and $0.640 \pm 0.002$, respectively. 
\begin{table*}[t]
    \caption{Relative populations for the NV$^-$ and NV$^0$ charge state and the determined Fermi energies. The errors represent the 95 $\%$ confidence interval.
    }
    \label{tab:NV_SiV_pop}
    \begin{tabular}{lllcccc}
       \hline 
       Index & Description & $[NV^{-}]_0$ & $[NV^{0}]_0$ & $E_F$ ($eV$) & $T_2$ ($\mu$s)\\
        \hline
        1 & E6 DNV14 ($[P1]=13$ ppm) & 0.56   $\pm$ 0.09   & 0.44 $\pm$ 0.09     & 2.62  $\pm$ 0.01 & 29 $\pm$ 5 \\
        2 & E6 DNV1  ($[P1]=0.8$ ppm) & 0.15   $\pm$ 0.02   & 0.85 $\pm$ 0.02     & 2.56  $\pm$ 0.01 &  52 $\pm$ 8 \\     
        3 & Sumitomo Ib ($[P1]>12.5$ ppm) & 0.89 $\pm$ 0.01  & 0.11 $\pm$ 0.01     & 2.66  $\pm$ 0.01  & 3.6 $\pm$ 0.4 \\
        4 & E6 TM220 & 0.18 $\pm$ 0.03 & 0.82 $\pm$ 0.03 & 2.57 $\pm$ 0.01    & 60 $\pm$ 10 \\
        5 & E6 TM220 & 0.29    $\pm$ 0.06   & 0.71    $\pm$ 0.06  & 2.58   $\pm$ 0.01 &  70 $\pm$ 20 \\
        6 & E6 TM220 & 0.07   $\pm$ 0.05   & 0.93   $\pm$ 0.05   & 2.54  $\pm$ 0.03  & 30 $\pm$ 10 \\
        7 & E6 TM200 & 0.57   $\pm$ 0.03   & 0.43   $\pm$ 0.03   & 2.62  $\pm$ 0.01  & 50 $\pm$ 9 \\
        8 & E6 TM200 & 0.6    $\pm$ 0.1    & 0.4    $\pm$ 0.1    & 2.62   $\pm$ 0.01 & 32 $\pm$ 4 \\
        \hline
    \end{tabular}
\end{table*}

\subsection{Fermi energy determination\label{sect:laser}}
Next, we study the dependence of the laser intensity on the NV charge state populations, aiming to extract the populations in the absence of laser excitation. For the study, we collected the PL spectra with various excitation laser powers, and then determined $[NV^{-}]$ and $[NV^{0}]$ using the PL decomposition analysis. Figure~\ref{fig:zerolaser}(a) shows the power dependence on  $[NV^{-}]$ and $[NV^{0}]$ for Sample 1. The results of the decomposition analysis for each laser power are provided in {\it Supporting Information} (SI) (see Sect. S1). 
As shown, the NV$^{-}$ (NV$^{0}$) relative population decreases (increases) as the excitation laser power increases. Therefore, the result shows the laser-induced charge state conversion from NV$^{-}$ to NV$^{0}$, which has been reported previously.~\cite{shields2015efficient, jayakumar2018spin}.
We further analyze the power dependence to extract the populations in the absence of the laser. 
Because the laser-dependent $[NV^{-}]$ and $[NV^{0}]$ exhibit the saturation behavior, we model the observed curve phenomenologically using the following stretched exponential function,
$y(x) = y_0+A\exp[-(x/\tau)^n]$
where $x$ is the variable, representing the laser power.
$A$, $\tau$ and $n$ are characteristic parameters.
In addition, the population in the absence of the laser excitation is given by $y(x = 0) = y_0 + A$.
Figure~\ref{fig:zerolaser}(a) shows the fit results using the stretched exponential function. 
The fits for both NV$^{-}$ and NV$^{0}$ data were done simultaneously with $[NV^{-}] = y(x)$ and $[NV^{0}] = 1-y(x)$. 
As can be seen, the agreement between the data and fits was excellent, with $R^2 =$ 97\% for Sample 1. 
We also obtained the $y(0)$ values from the fit results and also the populations in the absence of laser excitation ($[NV^{-}]_0$ and $[NV^{0}]_0$) to be $0.56 \pm 0.09$ and $0.44 \pm 0.09$, respectively. 
As shown in Fig.~\ref{fig:zerolaser}(b), $[NV^{-}]_0$ and $[NV^{0}]_0$ for Sample 4 were also obtained similarly.
The analyses of the other samples are also shown in Sect. S1 in SI.
As a summary, the relative populations for all samples are listed in Table~\ref{tab:NV_SiV_pop}.
The $[NV^{-}]_0$ and $[NV^{0}]_0$ values of the samples are also plotted in Fig.~\ref{fig:zerolaser}(c).
As shown, for Sample 1-3, the $[NV^{-}]_0$ values are larger when the concentrations of the P1 centers are larger. 
For Sample 4-8, the $[NV^{-}]_0$ values of the TM200 samples (Sample 7 and 8) are larger than those of the TM220 samples (Sample 4-6). 

Finally, the Fermi energy is determined. 
Using the results of $[NV^{-}]_0$ and $[NV^{0}]_0$ as well as Eqs.~\ref{eq:e_form} and \ref{eq:boltz}, we obtained the Fermi energies of the diamond samples. 
The results are summarized in Table~\ref{tab:NV_SiV_pop}. 
For Sample 1-3, the $E_F$ values vary from 2.56 to 2.66 $eV$.
Sample 3 has the highest Fermi energy among Sample 1-3.
This is because $E_F$ is higher when $[NV^{-}]_0$ is higher, as can be seen from Fig.~\ref{fig:formation}(c).  
For Sample 4-8, the $E_F$ values vary from 2.54 to 2.62 $eV$. 
Therefore, the $E_F$ range of the thermal-grade samples is slightly lower than that of the nitrogen-doped samples.

\section{Spin coherence and charge state stability of NV centers\label{sect:t2}}
\begin{figure}[t]
    \centering
     \includegraphics[width=13 cm]{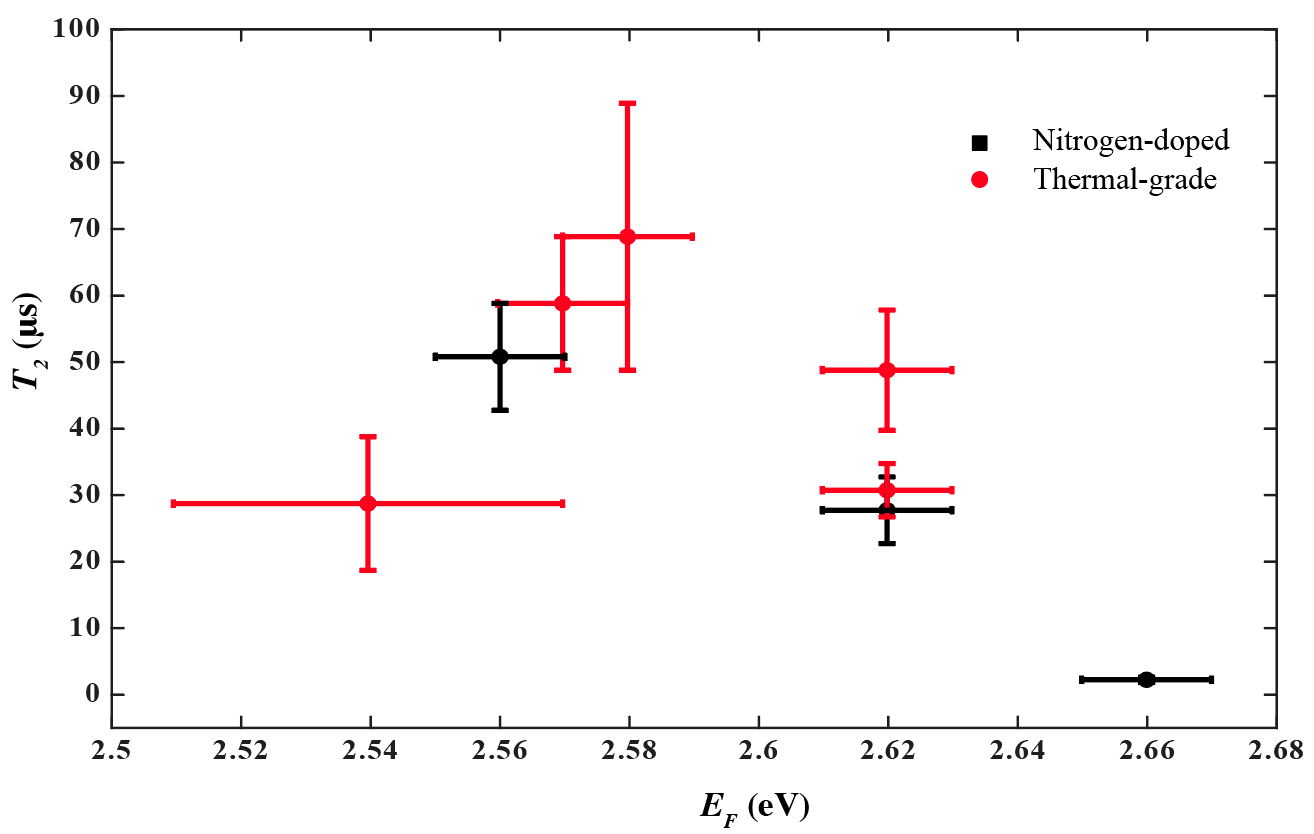}
    \caption{
    The Fermi energy ($E_F$) and $T_2$ of the nitrogen-doped and thermal-grade diamond samples. 
    }
    \label{fig:T2}   
\end{figure}
It is known that the spin decoherence time ($T_2$) in diamond depends on the concentration of the P1 centers. In particular, the concentration dependence on the $T_2$ time of the NV center and the P1 center has been studied extensively, and the studies showed that their $1/T_2$ and the P1 concentrations have a linear dependence~\cite{vanWyk_1997, ZHWang2013, Stepanov2016, Bauch2020}.
Here, we discuss the $E_F$ and the spin decoherence time ($T_2$).
In the study, we determined $T_2$ values for the NV center of the diamond samples using ODMR spectroscopy at zero fields~\cite{Abeywardana16, Fortman19}.
Table~\ref{tab:NV_SiV_pop} summarizes the results.
As can be seen, the $T_2$ values of the NV center for the nitrogen-doped samples (Sample 1-3) vary by more than an order of magnitude (3.6 to 52 $\mu$s) while the $T_2$ values of the thermal diamond samples (Sample 4-8) vary only in the range of $30-70$ $\mu$s.
Similar results were observed on $T_2$ of the P1 center studied using a 230 GHz electron paramagnetic resonance (EPR) spectrometer~\cite{Choi2014, Akiel2016} (see Sect. S3 and S4 in SI).
Figure~\ref{fig:T2} shows the $T_2$ and $E_F$ values together.
As can be seen, for the nitrogen-doped diamond samples (Sample 1-3), $T_2$ increases when $E_F$ decreases.
As we explained earlier, the Fermi energy ($E_F$) is determined by the concentrations of donors and acceptors, and the most common donor for diamond is nitrogen impurities.
Therefore, for the nitrogen-doped samples, $T_2$ clearly depends on $E_F$.
Moreover, we studied the thermal-grade diamond samples (Sample 4-8). 
Interestingly, the observed $T_2$ values for all thermal-grade diamond samples are long ($T_2 = 43-50$ $\mu$s), and the correlation between $E_F$ and $T_2$ is unclear (Fig.~\ref{fig:T2}).

Furthermore, the laser power dependence on the NV charge states was shown in~\ref{fig:zerolaser}(a) and (b).
As explained with Fig.~\ref{fig:zerolaser}(a), the laser power dependence of Sample 1 represents the laser-induced charge state conversion of the NV centers.
In particular, the $NV^-$ population decreases when the laser power increases; therefore, the result shows the conversion from $NV^-$ to $NV^0$.
Surprisingly, we observed opposite behavior in the thermal-grade diamond.
The laser power dependence for Sample 4 is shown in Fig.~\ref{fig:zerolaser}(b).
Unlike Sample 1, $[NV^{-}]$ increases with increasing laser power, resulting in the laser-induced charge state conversion from NV$^{0}$ to NV$^{-}$.
Therefore, the study shows that the laser excitation improves the stability of the NV$^{-}$ charge state.
The results from the other samples are also shown in Sect. S1 in SI. 
As shown, by increasing the excitation laser power,
$[NV^{-}]$ increases for all thermal-grade diamond samples (Sample 4-8) while $[NV^{-}]$ decreases for all nitrogen-doped diamond samples (Sample 1-3).
Therefore, the NV$^{-}$ centers in the thermal-grade diamond samples are more stable against laser excitation, and the NV charge state stability is unrelated to the $E_F$ value. 

These unique properties observed in the thermal-grade diamond indicate that the major donor in the system is not only nitrogen impurities but also other impurities, including the SiV centers, which were detected in the PL spectrum.
Consequently, these non-nitrogen impurities also influence the Fermi energies of the samples.
The result also implies that the impurities help the long coherence of the NV center and the stability against the charge state conversion.

\section{Summary}
In this work, we presented the method to determine the Fermi energy of diamond and its applications to investigate the $T_2$ and the stability of the NV centers. 
The determination method is based on the decomposition analysis of NV PL signals and the comparison with the DFT result.
Using the method, we demonstrated the determination of the Fermi energy of eight diamond crystal samples, including the nitrogen-doped and thermal-grade diamond samples.
We further investigated the relationship between the Fermi energy and $T_2$.
In this study, we unexpectedly found that thermal-grade diamond samples containing non-nitrogen donors are advantageous for achieving stable NV centers with long spin coherence.
Overall, our study highlights a PL method to measure the Fermi energy of diamond samples, which has advantages for measurements with fine spatial and fast temporal resolutions, and under harsh environments. 
The precision of the method may be improved by preparing NV PL standards from the sample itself for each decomposition analysis. 
The method can also be extended to other wide-bandgap semiconductors, including SiC, GaN, and hexagonal boron nitride (h-BN)~\cite{castelletto2020silicon, berhane2017bright, vaidya2023quantum}.

\section*{Acknowledgments}
This research was supported by the National Science Foundation (NSF) Award No. ECCS-2204667 (S.I. and S.B.C) and the Army Research Office (ARO) Award No. W911NF2210284 (S.I and S.B.C). Y.S., L.W., D.L., G.N., H.H., and S.T. also thank the support from NSF (ECCS-2204667).


\bibliography{ref}

\end{document}